\documentclass[aps,11pt,floatfix,twocolumn,tightenlines]{revtex4}
\usepackage[dvips]{color,graphicx}
\usepackage{bm}
\begin{document}
\newcommand{\boldsigma}{\mbox{\boldmath $\sigma$}}\newcommand{\boldxi}{\mbox{\boldmath $\xi$}}
\newcommand {\boldgamma}{\mbox{\boldmath$\gamma$}}
\newcommand{\boldtau}{\mbox{\boldmath $\tau$}}
\newcommand{\bftau}{\mbox{\boldmath $\tau$}}
\newcommand{\bfx}{{\bf x}}
\newcommand{\bfy}{{\bf y}}
\newcommand{\bfP}{{\bf P}}\def\bfS{{\bf S}}
\newcommand{\eq}[1]{Eq.~(\ref{#1})}\newcommand{\eqs}[2]{Eqs.~(\ref{#1},\ref{#2})}\def\poinc{Poincar\'{e} }
\def\bfq {{\bf q}}\newcommand{\bfxi}{\mbox{\boldmath $\xi$}}
\def\bfs {{\bf s}}\newcommand{\st}{{\scriptscriptstyle T}}\def\bfs {{\bf s}}
\def\bfn {{\bf n}}
\def\bfqp {{\bf q}_\perp}
\def\bfK{{\bf K}}
\def\bfKp{{\bf K}_\perp}
\def\bfL{{\bf L}}
\def\bfk{{\bf k}}
\def\bfp{{\bf p}}  
\newcommand{\bfkap}{\mbox{\boldmath $\kappa$}} 
\def\bfr{{\bf r}} 
\def\bfy{{\bf y}} 
\def\bfx{{\bf x}} 
\def\be{\begin{equation}}
 \def \ee{\end{equation}}
\def\bea{\begin{eqnarray}}
  \def\eea{\end{eqnarray}}
\newcommand{\eqn} {Eq.~(\ref )}
\newcommand{\bb}{\langle}
\newcommand{\kk}{\rangle}
\newcommand{\bk}[4]{\bb #1\,#2 \!\mid\! #3\,#4 \kk}
\newcommand{\kb}[4]{\mid\!#1\,#2 \!\mid}

\def\notp{{\not\! p}}
\def\notk{{\not\! k}}
\def\up{{\uparrow}}
\def\down{{\downarrow}}
\def\bfb{{\bf b}}
\setlength{\textheight}{9.20in}
\setlength{\textwidth}{7.6in}
\setlength{\topmargin}{-.840in}
\setlength{\oddsidemargin}{-.5125in}
\tolerance=1000

\def\poinc{Poincar\'{e} }
\def\bfq {{\bf q}}
\def\bfK{{\bf K}}
\def\bfL{{\bf L}}
\def\bfk{{\bf k}}
\def\bfp{{\bf p}}  
\def\be{\begin{equation}}
 \def \ee{\end{equation}}
\def\bea{\begin{eqnarray}}
  \def\eea{\end{eqnarray}}
\def\eqn {Eq.~(\ref )}

\newcommand{\kx}[2]{\mid\! #1\,#2 \kk}
\def\notp{{\not\! p}}
\def\notk{{\not\! k}}
\def\up{{\uparrow}}
\def\down{{\downarrow}}
\def\bfb{{\bf b}}

\title{
Non-Spherical Shapes of the Proton: Existence, Measurement and Computation }
\author{Gerald A. Miller}
\affiliation{ University of Washington
  Seattle, WA 98195-1560}


\sloppy



\maketitle

\noindent{\bf Introduction}
How can the spin 1/2 object known as the proton have a non-spherical shape? 
Why would a physicist even think of such a concept? Can a non-sphericity (or pretzelocity)
be measured or computed? This note is concerned with such questions.

The notion that the proton might not have  a spherical shape has its
 impetus  in the
  discovery that 
the spins of quarks and anti-quarks account for only about 30\% of the total angular momentum
\cite{oldspin}. Many 
 experiments have sought the origins of the remainder, expected to arise from quark and gluon
orbital angular momentum or from  pairs of strange quarks.

This article  is concerned with the relation between the quark orbital angular momentum
and the non-spherical shape of the proton. A number of concerns arise immediately.
 As a particle of spin
1/2, the proton can have no quadrupole moment, 
according to the Wigner-Eckart theorem.  In elastic electron-proton
scattering experiments,
the  effects of relativity  cause the initial and final 
wave functions to differ because their momenta differ.
For example, one  thinks  of a particle
in relativistic motion having a pancake shape because of the effects
of Lorentz contraction. Such an effect 
is  not a manifestation of the intrinsic  proton shape.

The presence of significant orbital angular momentum
can only  lead to a non-spherical shape if such can be defined by an
appropriate operator. 
 We used 
the  proton  model of
Ref.~\cite{Frank:1995pv}-\cite{Miller:2002ig}  to show  \cite{Miller:2003sa} 
  that the   rest-frame ground-state 
matrix elements of spin-dependent
density operators
reveal a host of non-spherical shapes.

\vskip.25cm
\noindent {\bf Experimental genesis}
The electromagnetic current matrix element can be written
in terms of the Dirac $F_1(Q^2)$ and Pauli $F_2(Q^2)$ form factors, 
$Q^2$ is the negative of the square of the space-like four-momentum transfer.
These form factors 
are probability amplitudes that
the proton can absorb a squared four momentum transfer $Q^2$ and still
remain  a proton. Two exist because the rapidly moving
quarks  within  the proton carry both
 charge and magnetization densities.
For $Q^2 = 0$ the form factors $F_1$ and $\kappa F_2$ are  the charge and the anomalous magnetic moment $\kappa$ in  units
 $e$ and
$e/2M_N$, and the magnetic moment  $\mu = 1+\kappa$.
The Sachs form factors are  
$
G_E = F_1 - {Q^2 \over 4M_N^2}\kappa F_2, 
G_M = F_1 +  \kappa F_2\;.
$ 

\begin{figure}
\unitlength1cm
\begin{picture}(5,5)(3,-6.0)
\includegraphics{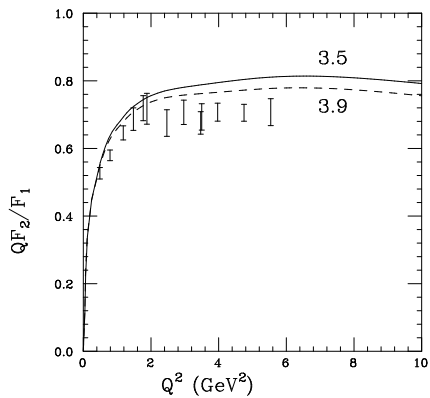}
\end{picture}
\caption{The ratio $QF_2/F_1$. 
The curves from the 1995 theory of \cite{Frank:1995pv}
for the ratio  are labeled by the
value of a  model parameter.
The  data 
are from 
\cite{Jones:1999rz} and 
\cite{Gayou:2001qd}. Figure reprinted with permission from 
\cite{Miller:2002qb}. Copyright 2002 by the American Physical Society.}
\label{ratio}
\end{figure}

In  the non-relativistic quark
model, $G_E$ and $G_M$ are  Fourier transforms
of the ground state  matrix elements of the  quark charge ($\sum_{i=1,3}
e_i\delta ({\bf r}-{\bf r_i})$ 
and magnetization $\sum_{i=1,3}
{e_i\over 2m_i}\delta ({\bf r}-{\bf r_i})$ density operators. Thus, non-relativistically, one expects  that  ${G_E(Q^2)/\mu G_M(Q^2)}=1.$ 
Interestingly,
the  opposite highly relativistic limit,  
in which dimensional counting applies, 
predicted \cite{Brodsky:1973kr} 
(using   the notion of  
helicity conservation in the interactions between  photons and
massless fermions)  that 
$\lim_{Q^2\to \infty} {QF_2/ F_1} ={m_q/ Q^2}, $ 
where $m_q$ is the small mass of a down or up quark. This is 
 equivalent to the non-relativistic expectation.
Thus  theoretical expectations (and  early data) were
that the ratio of the Sachs form factors would be  constant.
These expectations were dramatically thrown  aside with the discovery that $G_E/G_M$ falls rapidly with
increasing values of $Q^2$ and that  $QF_2/F_1$ is approximately constant\cite{Jones:1999rz},
 \cite{Gayou:2001qd}. See Fig.~\ref{ratio} which also displays the
results of our 1995 theory  \cite{Frank:1995pv,Miller:2002qb}.

An  explanation \cite{Miller:2002qb} of  how   the model of \cite{Frank:1995pv}
  describes the data 
 showed that the
constant 
ratio $QF_2/ F_1$ emerges from the model's  relativistic aspects. 
  For the proton wave function,
only the    component
in which the first two quarks 
have a vanishing total angular momentum  enters in computing the
electromagnetic form factors.
 Then the angular momentum  of the proton $S$  is governed  by that of the third quark.
The relevant Dirac spinor is:
\bea
 u(\bfK,S)={1\over \sqrt{E(K)+m_q}}\left(\begin{array}{c}
(E(K)+m_q)\vert S\kk \\
\boldsigma\cdot\bfK\vert S \kk 
\end{array}
\right),\label{spinor}\eea
with $E(K)=(K^2+m_q^2)^{1/2}$.
The magnetic quantum number of the proton is denoted by $S$, and
the lower component contains a term  $\boldsigma\cdot\bfK$ that allows the
quark to have a spin opposite to that 
of the proton's total angular momentum.
The  vector  $\bfK$ reveals the presence of the quark orbital angular momentum:
the struck quark
may carry a spin that is opposite to that of the proton. Consequently
nucleon  
helicity \cite{flip} is not conserved\cite{ralston,Braun:2001tj}.

\vskip0.25cm\noindent
{\bf Spin-dependent density operators}
We 
interpret  orbital angular momentum 
 in terms of the shapes of the 
proton by  these 
 are  exhibited 
through  the rest-frame ground-state 
matrix elements of spin-dependent
density operators \cite{Miller:2003sa}.
The usual  quantum mechanical density operator 
 is 
$ 
\widehat{\rho}(\bfr)= \sum_i 
\delta(\bfr-\bfr_i), $ 
where $\bfr_i$ is the position operator 
of the $i$'th particle; but
for  particles of  spin 1/2 one can measure
the {\it combined}  probability 
that particle is at a given position $\bfr$ and has a 
spin in an arbitrary, fixed direction specified by a unit vector
 $\bfn.$ The resulting   spin-dependent density SDD operator
is 
\bea \widehat{\rho}(\bfr,\bfn)= \sum_i
\delta(\bfr-\bfr_i){1\over2}(1+\boldsigma_i\cdot\bfn).\label{sddr}\eea 

To understand the connection between the  
 spin-dependent density and 
orbital angular momentum, 
  consider a first  example of a
 single charged particle
moving in a fixed, rotationally-invariant potential in an energy eigenstate
$|\Psi_{1,1,1/2,s}\rangle$ 
of quantum
numbers: $l=1,j=1/2$, polarized in the  direction $\widehat{\bfs}$
 and radial wave function $R(r_p)$. The wave function can be written
as 
$(\bfr_p| \Psi_{1,1,1/2,s}\rangle=R(r_p)\boldsigma\cdot\hat{\bfr}_p|s\rangle. $ 
The ordinary density. 
$\rho(r)=\langle\Psi_{1,1,1/2,s}|\delta(\bfr-\bfr_p)|
\Psi_{1,1,1/2,s}\rangle=R^2(r)$, a spherically symmetric result because  the 
effects of the Pauli spin operator square to unity. 
But 
the matrix element of 
the SDD 
is more interesting:
\bea \rho(\bfr,\bfn) 
={R^2(r)\over 2}\bb  
\widehat{s}\vert\boldsigma\cdot \hat{\bfr}(1+\boldsigma\cdot \hat{\bfn})\boldsigma\cdot\hat{\bfr}
\vert \widehat{s}\kk.\eea
The magnetic  quantum 
defines an axis, $\bfs$  
and
 the direction of vectors can be represented in terms of this axis:
$\hat{\bfs}\cdot\hat{\bfr}=\cos\theta$. Suppose  $\hat{\bfn}$
 is either parallel or anti-parallel to
the direction of the proton angular momentum  vector 
$\hat\bfs$. Then 
$ \rho(\bfr,{\bfn}=\hat{\bfs})={R^2(r)}\cos^2\theta,\;
 \rho(\bfr,{\bfn}=-\hat{\bfs}  )={R^2(r)}\sin^2\theta$, and
 the non-spherical shape is exhibited. 
The average of these
two cases   is  
a spherical shape. 

Another useful example is that of the 
Dirac  four-component  spinor  electron wave function of the 
 hydrogen
atom ground state, with relative size 
of the lower component governed by  the
fine structure constant, $\alpha$.  
The expectation value of the 
spin-dependent density operator, computed 
using  Dirac matrices, with $\boldsigma=\gamma^0\boldgamma\cdot\bfn$,
is 
$\rho(\bfr,\hat{\bfn}=\hat{\bfs})\propto
\left[1+\alpha^2/4\cos^2\theta\right] \sim1+10^{-5}\cos^2\theta$
with $\bfn=\hat{\bfs}$.
For $\bfn=-\hat{\bfs},\;\rho(\bfr,\hat{\bfn}=-\hat{\bfs})=
\alpha^2\sin^2\theta$/4.
Relativity as manifest by lower components of the  Dirac
wave function  causes the hydrogen atom to be  slightly, but definitely,  non-spherical! 

The notion of the SDD can be extended.
In condensed matter applications
 \cite{Prokes} 
neutrons interact with atomic electrons, and 
only the  (electronic) spin-dependent term of \eq{sddr}  
is used.
For quark systems,
the densities could be weighted by the charge or flavor 
of the quarks, or use  other operators.
 In particular, we use  \cite{Miller:2007ae} 
\bea
\widehat{\rho}_{\rm R}(\bfr,\bfn)\equiv \sum_i 
\delta(\bfr-\bfr_i){1\over2}(1+\gamma^0_i\boldsigma_i\cdot\bfn),\label{sddrrel}\eea 
which is more experimentally accessible.

\begin{figure}
\unitlength1.75cm
\begin{picture}(5,5)(1.70,.93)
\includegraphics{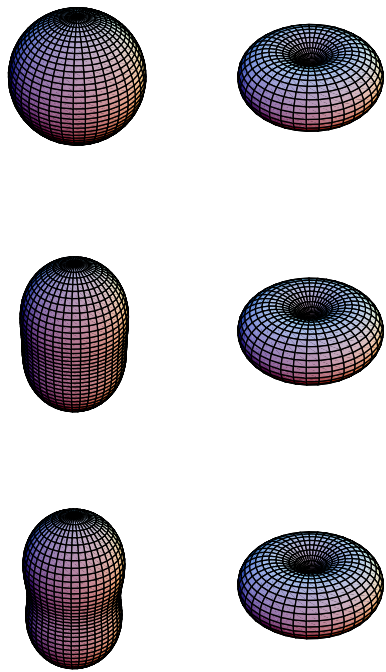}
\end{picture}
\label{fig:shapes}
\caption{ (Color online) 
Shapes of the proton. $\bfS$ is in the vertical direction.
Left column quark spin parallel to nucleon
spin. Right column : quark spin anti-parallel to nucleon spin. 
The value of $K$ increases from 0 to 1 to 4 GeV/c.  
Figure reprinted with permission from 
\cite{Miller:2003sa}. 
Copyright 2003 by the American Physical Society.} 
\end{figure}

Now  turn to the proton. Its  wave function
 is specified in momentum space,  so  we define \cite{Miller:2003sa} 
a  charge-weighted SDD  operator (the probability that  
a quark has 
a momentum
$\bfK$ 
and spin direction $\bfn$):
\bea\hat{\rho}(\bfK,{\bfn})=\int {d^3\xi\over(2\pi)^3} e^{i\bfK\cdot\boldxi}
\bar{\psi}(\boldxi){\widehat{Q}\over e}
(\gamma^0+\boldgamma\cdot\bfn\gamma_5)\psi({\bf 0}), 
\label{qft0}\eea 
where 
$\widehat{Q}/e$ is the quark charge operator in units of the proton charge.
The quark  field
operators are evaluated at equal time, $\xi^0=0$. For the case of 
$\widehat{\rho}_{\rm R}$ the term $\boldgamma\cdot\bfn$ is
replaced by $\gamma^0\boldgamma\cdot\bfn$.
For   a spin-polarized (in the $\hat{\bfS}$ direction)
proton at rest $\vert\Psi_{\bfS}\kk$, the matrix elements of 
$\hat{\rho},\hat{\rho}_{\rm R}$ (SDDs) are  given by
 \bea&&
\rho(\bfK,\bfn,\bfS) 
=A+B\bfn\cdot\hat\bfS+C
(\bfn\cdot\hat{\bfK}\;\hat\bfS\cdot\hat{\bfK} -{1\over3}\bfn\cdot\hat\bfS)\nonumber\\
&&\rho_{\rm R}(\bfK,\bfn,\bfS) 
=A_{\rm R}+B_{\rm R}\bfn\cdot\hat\bfS\nonumber\\&&+C_{\rm R}
(\bfn\cdot\hat{\bfK}\;\hat\bfS\cdot\hat{\bfK} -{1\over3}\bfn\cdot\hat\bfS)
\label{genshape}
,\eea
where $A,B,\cdots$ are scalar coefficients.
These forms represent the  most general {\it rest frame} 
shape of the proton, 
 if parity and rotational invariance are upheld \cite{Miller:2003sa}.

 We  display  the shapes of $\rho(\bfK,\bfn,\bfS)$ 
\cite{Miller:2003sa} in Fig.~2 
 for the cases of quark spin parallel and anti-parallel to
the polarization direction of the  proton $\bfS$. 
The shape for a given value of $K$ is determined by the ratio of the 
upper to lower components of the quark Dirac  spinor \eq{spinor}.
The relatively large value of the ratio 
 implies considerable non-sphericity and a sharp contrast between
the proton and hydrogen atom. As the value of $K$ increases from 0 to 4 GeV/c the shape varies from
that of a sphere to that of a peanut, if $\bfn\parallel\bfS$. 
The torus or bagel shape is obtained if  $-\bfn\parallel\bfS$. Taking $\bfn\perp\bfS$ leads to some very unusual shapes shown in Fig.~3. 
Using the given model \cite{Miller:2003sa}, one may also
 obtain in coordinate space SDDs. 
 Possible shapes  include a pretzel form \cite{Miller:2003sa}.


 Any 
wave function  yielding a non-zero value of the coefficient $C(\bfK^2)$ 
or $C_{\rm R}(\bfK^2)$
represents a system
of a non-spherical shape. If the relativistic constituent quark model 
of \cite{Frank:1995pv} is used, 
the  extra $\gamma^0$  changes the sign of the lower
component of the wave function, causing   
$C_{\rm R}=-C$.
Thus  either $C_{\rm R}$ or $C$ can be  used to infer information about
the possible shapes of the nucleon.
Measuring either 
 would require 
 controlling the   
{ three different vectors} $\bfn,\bfS$ and $\bfK$.

A specific  aspect of   $\hat{ \rho}(\bfK,\bfn)$
is easily related to completed experiments because
  $\int d^3K\hat{\rho}(\bfK,\bfn)$ 
 is a local operator. Its matrix element  is a linear combination of the
charge, integrals of spin-dependent structure 
functions $\Delta q$ (quark contribution 
to the proton total angular momentum), and $g_A$
that  can be determined  from previous  measurements.
We find
\bea&& \int d^3K\langle N\vert \hat{\rho}(\bfK,\bfn=\hat{\bfS},{\bfS})-\hat{\rho}(\bfK,\bfn=- \hat{\bfS},{\bfS})\vert N\rangle
\nonumber\\
&&={1\over6}(\Delta q+{1\over2}g_A)
= 0.68,\eea 
in which  $\Delta q=0.3$
 \cite{oldspin} and $g_A=1.26$.
The model we use gives $0.74$ for the above quantity, indicating
that shapes discussed here may not be unrealistic.


\vskip0.25cm {\bf Measuring the   Non-Spherical Shape of the Nucleon}
Can  non-spherical  shapes be measured? While measurements of the 
matrix elements of the non-relativistic spin-density operator 
\cite{Prokes}   reveal highly non-spherical densities, 
 finding the non-spherical nature of the proton 
has remained a challenge. 
Here we   explain  how
matrix elements of the spin-dependent density may be measured using their close connection with       
transverse momentum dependent 
parton densities. 
 
The densities of \eq{genshape} require  that
 the system be  probed with  
 identical  initial and final states. 
But this condition also 
 enters in measurements of  both ordinary and transverse-momentum-dependent
TMD parton distributions. 
The latter  
\cite{Mulders:1995dh} are:
\begin{widetext}
\begin{eqnarray}
\Phi^{[\Gamma]}(x,\bfK_T) & = &
\left. \int \frac{d\xi^-d^2\xi_\st}{2\,(2\pi)^3} 
\ e^{iK\cdot \xi}
\,\langle P,S \vert \overline \psi (0)\,\Gamma\,{\cal L}(0,\xi;n_-)
\,\psi(\xi) \vert P,S \rangle \right|_{\xi^+ = 0}, \label{projection}
\end{eqnarray}
\end{widetext}
where the specific 
path $n^-$ is  that of Appendix B of \cite{Mulders:1995dh}.
The functions $\Phi^{[\Gamma]}$ depend on the fractional momentum
$x$ = $K^+/P^+$, $\bfK_T$ and  on the hadron momentum
$P$. 
The operator $\Gamma$ can be any Dirac operator, {\it e.g.}
$\Gamma=i\sigma^{i+}\approx\sqrt{2} \gamma^0\gamma^i\gamma^5$, related to
$h_{1T}^\perp$,  which
causes the non-spherical nature of $\hat{\rho}_{\rm R}$.

\begin{figure}
\unitlength.81cm
\begin{picture}(5,5)(8.0,1.50)
\includegraphics{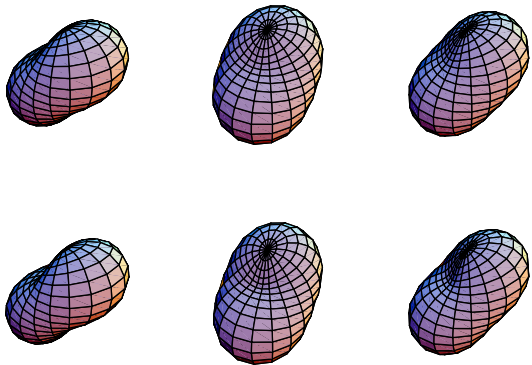}
\end{picture}\label{perp}
\smallskip
\caption{\label{fig:side}Shapes of the proton with
$\bfn\cdot \bfs=0$. 
Left column, $\bfn$ points  (out of page),
central: $\bfn$ points sideways,  right $\bfn$ is out of the page at a 45$^\circ$ angel.
 The momentum $K$ increases from
 1 to 4 GeV/c. Figure reprinted with permission from 
\cite{Miller:2003sa}. 
Copyright 2003 by the American Physical Society.}
\end{figure}

It is therefore tempting to 
try to associate an SDD  such as that  of \eq{qft0}
with TMDs, but one difference  is  essential. 
Parton density operators \eq{projection} 
depend on  quark-field operators defined
at a fixed  light cone time  $\xi^+=\xi^3+\xi^0=0$ while
our SDD 
is  an  equal-time, $\xi^0=0$, correlation 
function.
However, 
a relation
between the two sets of operators is obtained \cite{Miller:2007ae}
by 
  integrating  the TMD over all values of  $x$ setting $\xi^-$ to zero, 
and integrating \eq{qft0}
over all values of $K_z$ so that $\xi^3=0$. After integration, 
$\xi^\pm=0$ for both functions.
The density operators,
derived from those of \eq{genshape} are denoted by adding a $T$ to the subscript. 
Thus $\widehat{\rho}_{{\rm R}T}(\bfK_T,\bfn)\equiv\int_{-\infty}^\infty dK_z\widehat{\rho}_{\rm R}(\bfK,\bfn)$. It is therefore a matter of algebra to
show that
\bea &&
{\rho}_{{\rm R}T}(\bfK_T,\bfn_T,\bfS_T)/M=\tilde{f}_{1}(K_T^2)+\tilde{h}_{1}(K_T^2)\bfn\cdot\hat\bfS_T\nonumber\\&&
+{(\hat\bfn_T\cdot\bfK_T \hat\bfS_T\cdot\bfK_T-{1\over2}K_T^2\hat\bfn\cdot\hat\bfS_T)\over M^2}
\tilde{h}^\perp_{1T}(K_T^2),\label{rhotrel}
\eea
 in the rest frame, where a tilde is placed over each 
 TMD parton distribution to denote 
an $x$-integrated function.
Finding 
that non-zero value  of 
$\tilde{h}^\perp_{1T}\ne0$
 would demonstrate that the proton is not spherical.

The term  $\tilde{h}^\perp_{1T}$ 
causes  distinctive experimental signatures in 
semi-inclusive leptoproduction hadron production experiments  
 \cite{Boer:1997nt,Bacchetta:2000jk}. If the target is 
polarized in a  direction  $\bfS_T$ transverse to the 
lepton scattering plane, 
the cross
section acquires a term proportional to $\cos (3\phi_h^l)$ where
$\phi_h^l$  is the angle between the hadron production plane (defined by the momenta
of the incoming virtual photon and the outgoing  hadron) and the lepton scattering
plane. A similar effect occurs in electroweak semi-inclusive deep inelastic
leptoproduction 
\cite{Boer:1999uu}. 
In each of these cases, the momentum of the virtual photon and its
vector nature provide the analogue of  the  vector $\bfn$  
needed
to define the spin-dependent density. The hadronic  transverse momentum provides
the third, $\bfK_T$.
 Another  possibility occurs 
in the Drell-Yan reaction $pp(\uparrow)\rightarrow l\bar{l} X$,
 using one transversely polarized proton \cite{Boer:1999mm}.

The   shapes  inherent in  \eq{rhotrel} are illustrated using 
 the spectator model of \cite{Jakob:1997wg}.
Here
 $\phi$ is the angle between $\bfK_T$ and $\bfS_T$ and $\phi_n$ is the
angle between $\bfn$ and  $\bfS_T$. 
The transverse shapes of the proton (assuming a struck $u$ quark) 
are shown in Fig.~4, taking $\phi_n=\pi$. This
emphasizes the non-spherical nature
because the first two terms of \eq{rhotrel} tend to cancel. 
The  shapes of  \eq{rhotrel}
can be thought of as projections of the shapes displayed in previous figures.

\begin{figure}
\includegraphics[width=5cm]{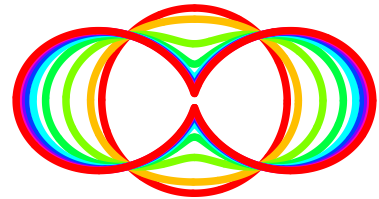}
\label{shape1} \caption{
Transverse shapes of the proton; 
$\sqrt{2}\hat\rho_{{\rm R}T}(\bfK_T,\bfn)/\tilde{f}_1(K_T^2)$. The horizontal axis is the the direction of $\bfS_T$ and $\bfn=\hat{\bfS}_T,\; 
\phi_n=\pi$. The shapes vary from  circular to highly deformed
as $K_T$ is increased from 0 to 2.0 GeV in steps of 0.25 GeV. 
Figure reprinted with permission from 
\cite{Miller:2007ae}. 
Copyright 2007 by the American Physical Society. }
\end{figure}
The model \cite{Jakob:1997wg} indicates that the
functions $f_1,h_1$ and $h_{1T}^\perp$ have very similar $x$ dependence, 
so that
measurements at values of $x$ for which these functions peak should be 
sufficient to construct  
 the required integrals over $x$.

The non-spherical nature of the nucleon shape is determined by  
the non-vanishing
of the TMD  $h_{1T}^\perp$. It is very exciting that experiments  planned  at 
Jefferson Laboratory aim to specifically measure $h_{1T}^\perp$ \cite{harut} and therefore
determine whether or not the  proton is  round.  

\vskip0.25cm\noindent{\bf Connection with lattice QCD}
The non-spherical shape of the nucleon can be established in lattice QCD by computing the lattice version of the 
angular integral of the
matrix element:
\bea
F_\Gamma(r)=\int d\hat{\bfr}Y_{20}(\hat{\bfr})
\bb\Psi_\bfS\vert\bar{\psi}(\bfr)
(\gamma^0+\Gamma\boldgamma\cdot\bfn\gamma_5)\psi({\bf 0})\vert\Psi_{\bfS}\kk
\nonumber\eea
where $\Gamma=1 $ or $\gamma^0$ and the link operator is not displayed.
 A non-zero value of $F(r)$ for any value of $r$  would immediately
tell us that the proton does not have a spherical shape. 
Matrix elements of  $\bar{\psi}_\alpha(\bfr)\psi_\beta({\bf 0})$ have been
evaluated for the case when the separation is one or two links. Thus the relevant information is available.
 Preliminary results for  $F_{\Gamma}(\bfr)$  exist only  for  separations of one-link, and 
current statistics 
are not high \cite{hwl}. Another possibility, closely related to finding
$h_{1T}^\perp$, 
 would be to take the spatial component of $\bfr$ to be perpendicular  to $\bfs$ and integrate over
the transverse directions. 
I hope  that the lattice QCD community will 
find it of sufficient interest to warrant the effort of a detailed, high-statistics calculation.

\vskip0.25cm\noindent
{\bf Summary}
The nature of the  proton wave function   
can be  elucidated by studying the matrix elements of a generalized density
operator.  Spin-dependent quark densities SDD are defined as
matrix elements of density operators in 
proton states of definite spin-polarization, and shown to have 
an infinite 
variety of non-spherical shapes. For high momentum quarks with spin parallel
to that of the proton, the shape resembles that of a peanut,
but  for quarks with anti-parallel spin the shape is that of a bagel.
 The matrix elements of the SDDs are closely
related to specific transverse momentum dependent TMD
parton distributions accessible in the angular dependence of
the semi-inclusive processes 
 $ep\rightarrow e\pi X$ and the Drell-Yan reaction
$pp\rightarrow l\bar{l}X$. New measurements or analyses would
 allow the direct exhibition of the non-spherical nature of the 
proton. The TMDs can be computed using lattice  QCD so that the non-spherical shapes could be measured experimentally and computed using fundamental theory.

\vskip0.25cm
\noindent {\bf Acknowledgments}
I thank the USDOE for partial support of this work. I thank 
C. Glasshauser
and J. Ralston for emphasizing
 the importance of understanding the shape of the proton and
  H. Avakian, D. Boer,  M. Burkardt,
W. Detmold, L. Gamberg, K. Hafidi, A.~Kvinikhidze,  J.W. Negele, 
 and J.C. Peng for useful discussions. 


\begin{thebibliography}{9}
\bibitem{flip} The ratio of $QF_2/F_1$ is a ratio of a 
 helicity-changing  to a helicity preserving matrix element.
\end{thebibliography}
\end{document}